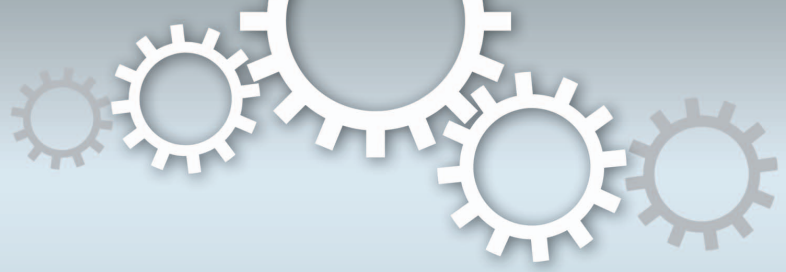

# SCIENTIFIC REPORTS

**OPEN**

# Magnetic Field Tunable Small-scale Mechanical Properties of Nickel Single Crystals Measured by Nanoindentation Technique

Hao Zhou, Yongmao Pei & Daining Fang

State Key Laboratory for Turbulence and Complex Systems, College of Engineering, Peking University, Beijing 100871, China.

Nano- and micromagnetic materials have been extensively employed in micro-functional devices. However, measuring small-scale mechanical and magnetomechanical properties is challenging, which restricts the design of new products and the performance of smart devices. A new magnetomechanical nanoindentation technique is developed and tested on a nickel single crystal in the absence and presence of a saturated magnetic field. Small-scale parameters such as Young's modulus, indentation hardness, and plastic index are dependent on the applied magnetic field, which differ greatly from their macroscale counterparts. Possible mechanisms that induced 31% increase in modulus and 7% reduction in hardness (i.e., the flexomagnetic effect and the interaction between dislocations and magnetic field, respectively) are analyzed and discussed. Results could be useful in the microminiaturization of applications, such as tunable mechanical resonators and magnetic field sensors.

Magnetic materials have been extensively employed in smart devices with micro- and nanofilms (e.g., memorizers, actuators, and sensors). However, the lack of feasible techniques induces difficulty in measuring small-scale mechanical and magnetomechanical properties. In the present study, a newly developed magnetomechanical nanoindentation technique is used to probe the small-scale magnetic field-dependent mechanical and coupling properties of a nickel single crystal.

For the past two decades, the depth-sensing nanoindentation technique[1], which measures the load and the displacement during indentation, has been extensively used to study the nano- and microscale mechanical properties of structural[2], functional[3], and biological[4] materials. Compared with their macro-scale counterparts, small-scale properties of materials usually show arresting anomalism[5,6], which has recently gained considerable attention. Regulation and control of these properties by changing the environmental variables have become widely studied topics in materials science. For example, the effects of temperature[7], electric current[8], water environment[9], and interface misfit strain[10] on small-scale mechanical properties have been investigated using the nanoindentation technique. However, only a few studies have reported on the effect of magnetic field on these properties, although much work has been undertaken on the macroscale properties. For example, the magnetic field-dependent Young's moduli of magnetostrictive materials have been measured using different methods such as quasi-static tensile/compression[11], vibrating reed method[12], ultrasonic wave velocity[13], magnetomechanical resonance[14], and optical method[15]. The magnetic field-dependent yield strength and fatigue life of materials have been analyzed with uniaxial tensile and four-point bending methods[16]. We attempt to determine the effect of magnetic field on the small-scale mechanical and magnetomechanical properties.

In this study, we develop a new magnetomechanical nanoindentation technique and conduct nanoindentation tests on a nickel single crystal with and without a saturated magnetic field to probe the magnetic field dependency of small-scale mechanical properties. The magnetic field alters these properties such as the Young's modulus, indentation hardness, yield strength, plastic index, energy dissipation ratio, and resonant frequency. Possible mechanisms for the field-induced increase in Young's modulus and reduction in hardness are analyzed and discussed. An empirical characterization method is established to determine the magnetomechanical parameters, and yield the local magnetomechanical coupling coefficient. These results can be used for the design and miniaturization of applications such as tunable mechanical resonators[17], magnetic field sensors[18], active vibration control systems[19], magnetic force control systems[20], tribology control systems[21], and graphene growth[22].

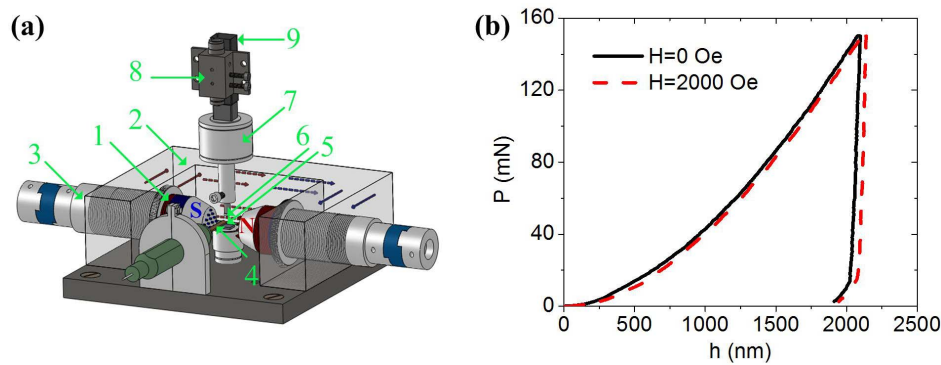

**Figure 1** | (a) Schematic of the magnetomechanical nanoindentation apparatus. 1. Permanent magnets with blue and red parts representing S and N poles, respectively. 2. U-shaped magnet yoke made of electrical pure iron to conduct the magnetic flow. 3. Mechanical transmission parts to adjust the distance between the magnetic poles via rotation. 4. Hall effect magnetometer to measure the magnetic field strength. 5. Sample. 6. Indenter probe. 7. Load sensor. 8. Displacement sensor. 9. Piezoelectric actuator. (b) Nanoindentation curves on Ni(111) sample with and without a horizontal saturated magnetic field.

## Results

**Design of the magnetomechanical nanoindentation apparatus.** Figure 1a shows the new magnetomechanical nanoindentation apparatus comprising a specially designed magnetic field device made of rare-earth permanent magnets (e.g., NdFeB) and pure iron, mechanical loading part, and measurement system. Changing the distance between two poles tunes the magnetic field strength in a sample region. All parts of the apparatus exposed to the magnetic field are fabricated from nonmagnetic materials to avoid the interference between the magnetic field and other apparatus components. Magnetic field-independent tests on standard sample (fused quartz) indicate the feasibility of this apparatus. The details of the design and calibration of the apparatus are listed on Ref. 23.

**Characterization of small-scale mechanical and magnetomechanical parameters.** Indentations are conducted on Ni(111) single crystal samples with and without an external horizontal magnetic field of 2000 Oe with this apparatus. This magnetic field, which is stronger than the saturated magnetic field strength of 1400 Oe, is used to ensure the magnetic saturation state[24]. Indentations with the magnetic field are first conducted, and then those without the magnetic field. Given the soft magnetic properties (low coercivity and residual magnetization) of nickel, the effect of magnetic hysteresis on the $\Delta E$ values is small[25] and is not addressed in this study. Figure 1b shows the typical indentation load-depth. Fifty to sixty indentations with and without magnetic field are conducted for the statistical averaging to minimize the experimental error. The maximum depth increases by 2.67% on the average upon the application of the magnetic field. Adopting the widely used Oliver-Pharr method[1], the upper portion of the unloading curves is approximated by the power law relation: $P = \alpha(h - h_f)^m$, where $P$ and $h$ refer to the load and depth at any time of the indentation, and $h_f$ is the residual depth, which increases by 4.14% in the presence of magnetic field. $\alpha$ and $m$ are the fitting constants. The field-induced increase in the power law exponent is small (0.56%). The initial unloading stiffness is obtained by calculating the derivative of the load with respect to the depth at the maximum depth, $S = (dP/dh)|_{h=h_{max}}$, which increases significantly by 30.35% on the average. The effective and Young's moduli can be obtained using $E_{eff} = \frac{\sqrt{\pi}}{2} \frac{S}{\sqrt{A}}$ and $\frac{1}{E_{eff}} = \frac{1-\nu^2}{E} + \frac{1-\nu_i^2}{E_i}$, where $E$ and $\nu$ are Young's modulus and Poisson's ratio of the sample, respectively. $E_i$ and $\nu_i$ are Young's modulus and Poisson's ratio of the indenter probe, respectively, and $A$ is the projected area of the contact. The indentation hardness is defined as the maximum load divided by the contact area, $H_{IT} = P_{max}/A$. Figure 2 shows the results of Young's modulus and indentation hardness. The saturated field increases the value of $E$ by 31.36% (from 160.29 ± 36.22 to 210.55 ± 40.97), but decreases the value of $H_{IT}$ by 6.77% (from 1.56 ± 0.16 to 1.45 ± 0.15). Several other properties of materials or structures, such as beams, can be determined based on the modulus and hardness as follows:

1) Hardness-to-modulus ratio $H_{IT}/E_{eff}$. In tribology, the wear resistance of hard coatings is better described as $H_{IT}/E_{eff}$, instead of hardness alone[26]. In fracture mechanics, this ratio is used to determine the fracture toughness[27].

2) Energy dissipation ratio $W_p/W_{tot}$. Integrating the loading and unloading curves yields the total work $W_{tot}$ and the reversible work, $W_u$, respectively. Then, the irreversible work is $W_p = W_{tot} - W_u$, and the energy dissipation ratio is $W_p/W_{tot}$. $W_p/W_{tot} = 0$ for a purely elastic indentation, and $W_p/W_{tot} = 1$ for a purely plastic indentation.

3) Yield strength $Y$. Based on the dimensional analysis and finite element calculations for the conical indentation[28], the yield strength of elastic-plastic solids can be obtained via the approximate relationship: $Y = \begin{cases} H_{IT}/2.8, & \text{for } Y/E \to 0.0 \\ H_{IT}/1.7, & \text{for } Y/E \to 0.1 \end{cases}$.

4) Plasticity index $\psi$. In tribology, the plasticity index describes the deformation properties of a rough surface in contact with a smooth surface[29], which is expressed as $\psi = (E_{eff}/H_{IT})\sqrt{A/R}$, where $\sqrt{A/R}$ is a geometric factor. The deformations are predominantly elastic or plastic when $\psi$ is much less than or greater than unity, respectively.

5) Resonant frequency of structures $f_0$. In structural mechanics, the resonant frequency of structures (e.g., rods and beams) is a function of the Young's modulus. For example, the resonant frequency of a cantilever film or beam is proportional to the square root of Young's modulus, i.e., $f_0 \propto \sqrt{E(H)}$. A magnetic field-dependent Young's modulus regulates the resonant frequency of nano- and microstructures.

6) Magnetomechanical coupling coefficient $k$. This coefficient is defined as the ratio of the output mechanical energy to the input magnetic energy[30] and can be expressed as $k = (1 - E_0/E_s)^{1/2}$, where $E_0$ and $E_s$ are the moduli of the sample in the absence and presence of the saturated magnetic field, respectively. A smaller $E_0/E_s$ implies larger $k$ and higher energy conversion efficiency.

Table 1 shows the small-scale properties determined using the above equations in this paper. These properties (Young's modulus $E$, indentation hardness $H_{IT}$, yield strength $Y$, plastic index $\psi$, and the resonant frequency of microcantilever beam $f_0$) change remarkably with the saturated magnetic field. The local magnetomechanical coupling coefficient of $k = 0.49$ is obtained, which is higher than that at the macroscale[15] (from 0.13 to 0.17). The ratio of indentation hardness to effective modulus $H_{IT}/E_{eff}$ and the ratio of the irreversible work to the total work of the indentation process $W_p/W_{tot}$ are also analyzed because these parameters are important in tribology and fracture mechanics[26–29].

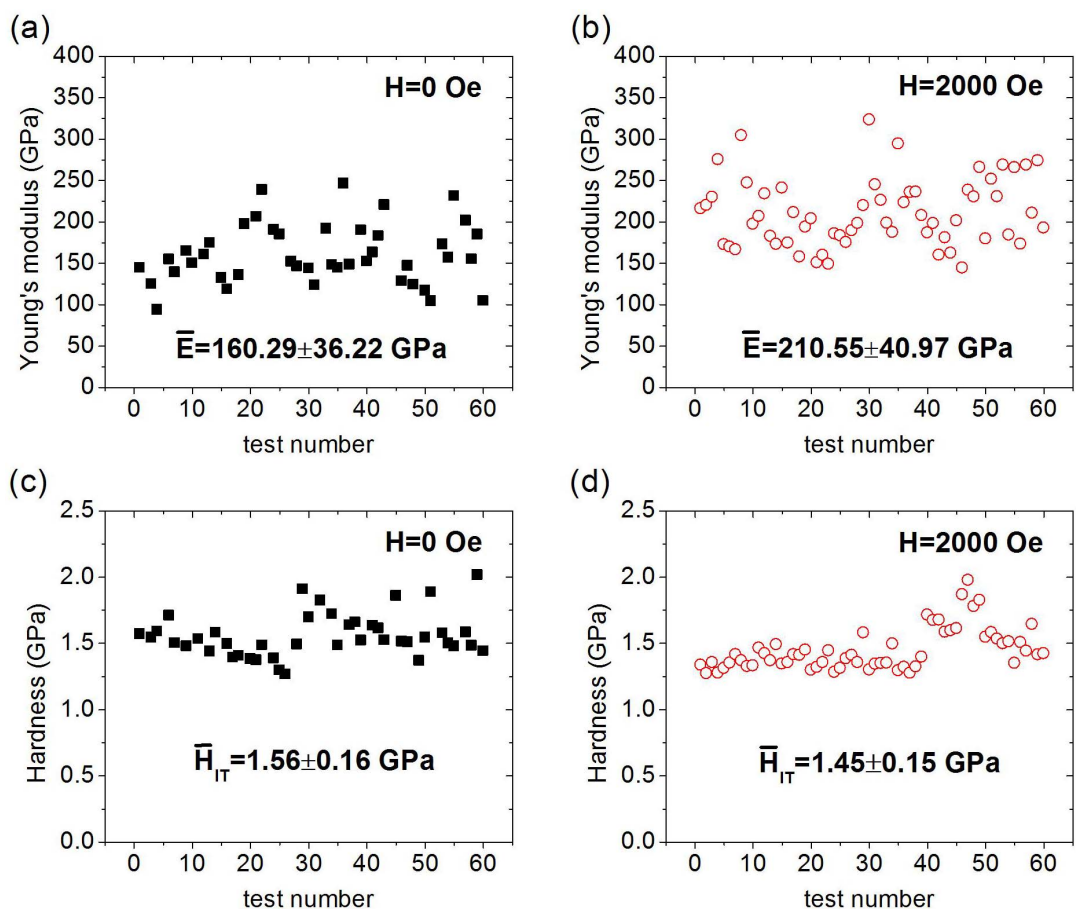

Figure 2 | (a and b) Young's moduli and (c and d) indentation hardnesses of the Ni(111) sample with and without a saturated magnetic field.

**Empirical relationship between energy dissipation and hardness-to-modulus ratios with and without a saturated magnetic field.** We briefly introduce the empirical relationship between energy dissipation and hardness-to-modulus ratios determined via dimensional analysis and finite element calculations, $H_{IT}/E_{eff} = f(W_p/W_{tot})$, which was proposed by Cheng and Cheng[31]. The relationship is based on the assumption of elastic-plastic materials with uniaxial constitutive equations given as $\sigma = \begin{cases} E\varepsilon, & \text{for } \varepsilon \leq Y/E \\ K\varepsilon^n, & \text{for } \varepsilon \geq Y/E \end{cases}$, where $\sigma$ and $\varepsilon$ are the stress and strain, respectively, $K$ is the strength coefficient, and $n$ is the work-hardening exponent. The solid is elastic below the yield strength $Y$. $E$ and $\nu$ are used to characterize the deformation. The solid is work-hardening after yielding.

Consequently, $E$, $Y$, $n$, and $\nu$ are sufficient to describe the stress-strain relationship. Figure 3 plots the finite element calculations (hollow squares) according to Ref. 31, which correspond to elastic-plastic materials with $Y/E = 0$–$0.006$, $\nu = 0.2$–$0.4$, and $n = 0$–$0.5$. Fitting the calculation points yields an empirical equation

$$H_{IT}/E_{eff} = -0.166 \times W_p/W_{tot} + 0.166 \quad (1)$$

Published experimental data for three common metals (W, Cu, and Al as blue solid dots from left to right in Figure 3) are also presented[31], which show consistency with the fitting line. However, the validity for ferromagnetic materials with strong magnetomechanical coupling properties is yet to be examined. In the present study, the experimental data of the ferromagnetic nickel single crystal in the absence and presence of an external saturated magnetic field are plotted (red five-pointed stars from left to right in Figure 3), with the error bars indicating the standard deviations. The saturated field increases $W_p/W_{tot}$ by 2.50% (from $0.92 \pm 0.02$ to $0.95 \pm 0.01$), but decreases $H_{IT}/E_{eff}$ by 27.36% (from $10.68 \pm 2.57$ to $7.76 \pm 1.48$, unit: $10^{-3}$). Results with and without external saturated field agree with the empirical equation. Thus, a traditional dimensionless relationship for elastic-plastic materials to the magnetomechanical coupled dimensionless relationship for ferromagnetic materials is feasible. The deformation and energy dissipation behavior of nickel can be regulated in line with this relationship by varying the external magnetic field, which is a noncontact method.

| Table 1 | Properties of nickel sample in the absence/presence of a saturated magnetic field | | | | | | | |
|---|---|---|---|---|---|---|---|---|
| | $E$ [GPa] | $H_{IT}$ [GPa] | $H_{IT}/E_{eff}$ [$10^{-3}$] | $W_p/W_{tot}$ | $Y$ [GPa] | $\psi$ | $f_0$ [Hz] | $K$ |
| H = 0 Oe | 160.29 ± 36.22 | 1.56 ± 0.16 | 10.68 ± 2.57 | 0.92 ± 0.02 | 0.56 ± 0.06 | \ | \ | 0.49 |
| H = 2000 Oe | 210.55 ± 40.97 | 1.45 ± 0.15 | 7.76 ± 1.48 | 0.95 ± 0.01 | 0.52 ± 0.06 | \ | \ | \ |
| Average Variation | 31.36% | −6.77% | −27.36% | 2.50% | −6.77% | 34.66% | 14.80% | \ |

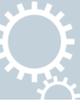

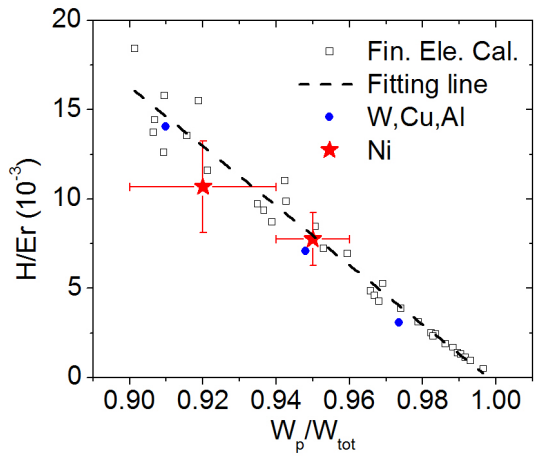

Figure 3 | Empirical relationship between $W_p/W_{tot}$ and $H_{IT}/E_{eff}$.

Table 1 shows that the magnetic field changes all the eight properties of nickel. However, only $E$ and $H_{IT}$ are independent. The following reasons for the magnetic field dependence are proposed.

**Magnetic field-induced increase in the indentation elastic modulus.** The variation in $E$ between the saturated and demagnetized states is called the $\Delta E$ effect given as $\Delta E/E_0 = (E_s - E_0)/E_0$, where $E_s$ and $E_0$ are the Young's moduli for saturated and demagnetized states[12–15], which can be explained from the perspective of the micromagnetic energies. On the one hand, ferromagnetic materials possess spontaneous magnetization because of the strong exchange interaction between the spins below the Curie temperature. Positive exchange interaction yields parallel arrangement of the magnetic moments of the atoms and creates a single domain state. On the other hand, the occurrence of a multi-domain state should reduce the magnetostatic energy of the whole sample. The magnetization of the domains orients in the directions of the easy axes to decrease the anisotropic energy. The orientation of the domains with the spontaneous strain will also change by an applied stress field because of the mechanical anisotropy that elongates or contracts the sample in certain directions. The total reversible strain comprise the purely elastic and domain rotation strains, in which $\varepsilon_{total} = \varepsilon_{elast} + \varepsilon_{domain}$. All the magnetic moments arrange in the same direction as the external magnetic field in the case of a highly saturated state to decrease the external magnetic energy. The mechanical disturbance cannot induce domain rotation, so that no domain rotation strain exists upon application of mechanical loading. Here, the reversible deformation is purely elastic, $\varepsilon_{total} = \varepsilon_{elast}$. Therefore, the Young's moduli of the demagnetized and saturated states can be expressed as $E_0 = \sigma/(\varepsilon_{elast} + \varepsilon_{domain})$ and $E_s = \sigma/\varepsilon_{elast}$, respectively. Hence,

$$\Delta E/E_0 = (E_s - E_0)/E_0 = \varepsilon_{domain}/\varepsilon_{elast} \quad (2)$$

The domain rotation strain at the macroscale is attributed to the converse magnetostrictive effect, which describes the stress-induced rotations of the magnetic domains:

$$\varepsilon_{domain} = \varepsilon_\lambda \quad (3)$$

However, the stress gradients can be much larger at the nano/microscale than that at the macroscopic scale because of the existence of small-scale inhomogeneities, dislocations, or strain relaxations[32–34]. The flexomagnetic or flexoelectric effect, a stress gradient-induced magnetization or polarization, has an important function. Therefore, the proposed macroscale model should be modified at small scale. The stress/strain gradient or the breakage of crystal lattice symmetry changes the magnitude and the orientation of spontaneous magnetization[35] as well as the spontaneous strain of each domain that can contribute to $\varepsilon_{domain}$ through the domain rotations. We consider the flexomagnetic effect to modify the domain rotation strain as

$$\varepsilon^I_{domain} = \varepsilon^I_\lambda + \varepsilon^I_\mu \quad (4)$$

The second term on the right side of Equation (4) describes the flexomagnetic strain. The superscript $I$ indicates the small-scale parameters in the nanoindentation model. The stress field inside the sample induced by the nanoindentation is inhomogeneous. For simplification, we introduce the representative stress $\sigma^I$ and representative strain $\varepsilon^I$ and rewrite the expression for the modulus determined from the Oliver-Pharr method in the form of the uniaxial stress state by $E^I_0 = \sigma^I/(\varepsilon^I_{elast} + \varepsilon^I_\gamma + \varepsilon^I_\mu)$ and $E^I_S = \sigma^I/\varepsilon^I_{elast}$ for demagnetized and saturated states, respectively. Then, the $\Delta E$ effect in the nanoindentation model can be expressed as

$$\Delta E^I/E^I_0 = (E^I_S - E^I_0)/E^I_0 = (\varepsilon^I_\gamma + \varepsilon^I_\mu)/\varepsilon^I_{elast} \quad (5)$$

Table 1 shows that the average Young's modulus for the saturated state is 210.55 GPa, which is in accordance with published values from 191.5 GPa to 228.5 GPa[36]. By contrast, the average Young's modulus for the demagnetized state is only 160.29 GPa, which is obviously smaller than the reported value of 170 GPa to 231 GPa[36]. Consequently, the small-scale value $\Delta E^I/E^I_0 = 31\%$ is larger than the macroscale value (from 1% to 20%), which depends on the grain size or the internal stress. Larger the grain sizes imply larger $\Delta E$. We use the maximum value of 20% at the macroscale for a nickel single crystal given that the samples possess various grain sizes provided in previous studies. Thus, the value of the flexomagnetic contribution to the small-scale $\Delta E$ effect can be tentatively estimated. The stress/strain gradient field inside the samples at the macroscopic experiments is negligible compared with that at small-scale. The ratio of converse magnetostrictive strain to purely elastic strain at both macro- and small-scale is approximately given as

$$\varepsilon^I_\gamma / \varepsilon^I_{elast} \approx \varepsilon_\lambda / \varepsilon_{elast} \quad (6)$$

Equations (2) to (6) yield the ratio of the flexomagnetic strain to the converse magnetostrictive strain:

$$\varepsilon^I_\mu / \varepsilon^I_\gamma = \frac{\Delta E^I}{E^I_0} \bigg/ \frac{\Delta E}{E_0} - 1 \approx 0.5 \quad (7)$$

Statistically, the strain resulting from the flexomagnetic effect can be as much as 50% of the strain from the converse magnetostrictive effect on the average.

Figure 4 shows a schematic of the domain switches during unloading. The red arrows indicate the magnetization orientation of the domains, and the long axis direction of blue ellipses is the spontaneous strain direction. The spontaneous strain is particular to the spontaneous magnetization because nickel is a negative magnetostrictive material. Eight <111> easy magnetization directions in the single crystal nickel exist theoretically, resulting in the domain walls of 180°, 109.47°, and 70.53°. Only two assumed easy axes are plotted in Figure 4 with domain walls of 180° and 90° as a two-dimensional simplification. Branched structures of the volume domains should exist near the surface[24], but are neglected in this study to simplify the analysis. The unloading deformation of the demagnetized state comprised a purely elastic part and a reversible domain rotation part, whereas that of the saturated state is purely elastic. Thus, the reversible displacement of the deformed surface around the indenter tip in the saturated state is smaller than that in the demagnetized state because of the absence of the reversible domain rotation strain. Figure 4a indicates the indentation-induced magnetic anisotropy by the net magnetization in the perpendicular direction. Sort et al.[37] found the indentation-induced perpendicular magnetic anisotropy and the enhanced magnetic properties in ferromagnetic metallic glass in the absence of an external magnetic field.

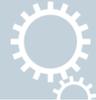

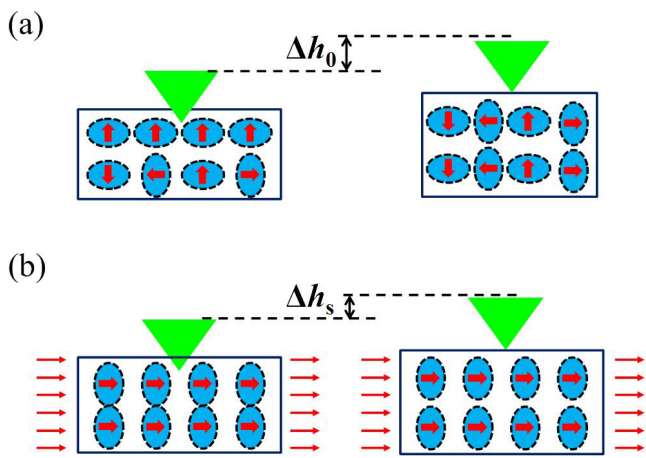

**Figure 4 | Schematic of the unloading behavior of nanoindentation processes with and without the saturated magnetic field.** The reversible deformation $\Delta h_s < \Delta h_0$ caused by the saturated magnetic field inhibits the deformation from domain rotations.

**Magnetic field-induced reduction in indentation hardness.** Figure 2 and Table 1 show that the magnetic field reduces the indentation hardness and yield stress by 6.77%. The magnetoplastic effect should be related to the dislocation generation, movement, and interactions. From the perspective of energy, the dislocations require extra energy to pass through the magnetic domain walls in the absence of a saturated magnetic field. By contrast, most of the domain walls vanish and the mobility of dislocations increases in the presence of a saturated magnetic field. Therefore, plastic deformation becomes easier in the saturated state. This observation is similar to the report of Bose on the magnetic field-dependent fatigue life of steel[16]. The dislocation structures of niobium after 5% uniform elongation with and without 1500 Oe magnetic field are observed with a transmission electron microscope[38]. The magnetic field increases the mobility and decreases the reactions of dislocations, which provides a direct evidence for the interaction between the magnetic field and the dislocations. The variation in the magnetic ordering, which can interact directly with the external magnetic field, in the dislocation region is also possible[39]. The plasticity of materials is intensified from the interactions between the magnetic field and the dislocations. Therefore, the hardness and the yield stress decrease, and the energy dissipation ratio increases.

## Discussion

Nano- and micromagnetic materials are widely used in functional devices. However, the measurement of small-scale mechanical and magnetomechanical properties is still challenging, which restricts both the design of new products and the performance of smart devices. We develop a new magnetomechanical nanoindentation technique, in which the mechanical properties, such as Young's modulus (up to 31%) and hardness (up to −7%), are magnetic field-tunable. However, the results differ from those of the macro-scale counterparts. The flexomagnetic effect and the interaction between dislocations and magnetic field are proposed to explain these phenomena. Like that temperature (cooling) can induce symmetry breaking via phase transition, stress/strain gradient resulting from surface/interface, defects, bending/torsion, contact stress concentration, or nanoindentation can also change the local symmetry and contribute to the magnetization of materials[40–42]. Proofs for these phenomena have been found in ferromagnetic and antiferromagnetic materials such as cobalt[43], nickel[44], and $BiFeO_3$[6]. Nanoindentation technique has been employed to study the flexoelectric effect in ferroelectric materials, which is similar to our study[45,46]. The interaction between the dislocations and the magnetic field has been experimentally verified in niobium[38].

Results from this study could be useful in microminiaturization of applications such as tunable mechanical resonators and magnetic field sensors. The magnetic field could control the mechanical, tribological, and structural properties of these devices. However, we encourage additional experiments or calculations to ascertain the feasibility of our proposal for the mechanism, considering that the analyses may not provide a full insight into the effects of the magnetic field on the small-scale mechanical properties. In situ observation via advanced microscopy technique is one of the directions of future research.

## Methods

**Sample information.** The nickel single crystal is provided by MTI Corporation in Hefei, China, and is grown using Bridgman method. The thickness of this sample is approximately 200 μm, and surface roughness of the polished side is less than 0.5 nm. This side is used for nanoindentation measurements.

**Magnetomechanical nanoindentation experiments.** The mechanical module of the nanoindentation apparatus has maximum load of 1 N, load noise of 35 μN, maximum depth of 20 μm, and depth noise of 8 nm. The performance details of the apparatus are available from Ref. 23. A three-sided Berkovich diamond indenter probe with tip radius of approximately 150 nm is used to penetrate the sample. Each indentation test is performed within 120 s, including 30 s holding time at the peak load to ensure the sufficient loading deformation of nickel from the indentation creep effect. The testing temperature is kept at 23 ± 1°C to reduce the thermal drift. Fifty to sixty indentations with and without magnetic field are conducted for statistical averaging to minimize the experimental error. A Gauss meter (Beijing CUIHAIJIACHENG Magnetic Technology Company, CH-1300), which range from 0 kOe to 30 kOe with a resolution of 1 Oe, is used to measure the magnetic field strength. The Oliver-Pharr method is adopted to calculate the Young's modulus and the indentation hardness from the experimental curves. The upper 30% of the unloading curves are fitted with the power law relation: $P = \alpha(h - h_f)^m$, where $h_f$ is the residual depth, and $\alpha$ and $m$ are the fitting constants.

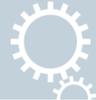

### Acknowledgments

The authors are grateful for the support by the National Natural Science Foundation of China (Grant Nos. 11090330 and 11090331) and the Chinese National Programs for Scientific Instruments Research and Development (Grant No. 2012YQ03007502). The support by the National Basic Research Program of China (Grant No. G2010CB832701) is also acknowledged.

### Author contributions

H.Z. performed the nanoindentation experiments and analysis. Y.P. and D.F. supervised the project. H.Z., Y.P. and D.F. discussed the equipment design, experimental scheme, and experimental results. All authors contributed to and discussed the manuscript.

### Additional information

**Competing financial interests:** The authors declare no competing financial interests.

**How to cite this article:** Zhou, H., Pei, Y.M. & Fang, D.N. Magnetic Field Tunable Small-scale Mechanical Properties of Nickel Single Crystals Measured by Nanoindentation Technique. *Sci. Rep.* **4**, 4583; DOI:10.1038/srep04583 (2014).